\documentclass[12pt]{iopart}
\usepackage{iopams}

\usepackage{graphicx}
\usepackage[utf8]{inputenc}
\usepackage[T1]{fontenc}
\usepackage{dcolumn}
\usepackage{bm}
\usepackage{iopams} 
\usepackage{color}
\usepackage{soul}
\usepackage{url}
\usepackage[normalem]{ulem}
\usepackage{aas_macros}

\usepackage[caption=false]{subfig}

\begin{document}

\title[CNN classifier for the distributions of $\mathcal{F}$-statistic signal candidates]{Convolutional neural network classifier for the output of the time-domain $\mathcal{F}$-statistic all-sky search for continuous gravitational waves}

\author{Filip Morawski}
\address{Nicolaus Copernicus Astronomical Center, Polish Academy of Sciences, Bartycka 18, 00-716, Warsaw, Poland}
\ead{fmorawski@camk.edu.pl}

\author{Micha{\l} Bejger}
\address{Nicolaus Copernicus Astronomical Center, Polish Academy of Sciences, Bartycka 18, 00-716, Warsaw, Poland}

\author{Pawe{\l} Cieciel{\k a}g}
\address{Nicolaus Copernicus Astronomical Center, Polish Academy of Sciences, Bartycka 18, 00-716, Warsaw, Poland}

\begin{abstract}
Among astrophysical sources in the Advanced LIGO and Advanced Virgo detectors'
frequency band are rotating non-axisymmetric neutron stars emitting
long-lasting, almost-monochromatic gravitational waves. Searches for these
continuous gravitational-wave signals are usually performed in long stretches
of data in a matched-filter framework e.g., the $\mathcal{F}$-statistic method.
In an all-sky search for {\it a priori} unknown sources, large number of templates is
matched against the data using a pre-defined grid of variables (the
gravitational-wave frequency and its derivatives, sky coordinates),
subsequently producing a collection of {\it candidate signals}, corresponding
to the grid points at which the signal reaches a pre-defined signal-to-noise
threshold. An astrophysical signature of the signal is encoded in the
multi-dimensional vector {\it distribution} of the candidate signals. In the first
work of this kind, we apply a deep learning approach to classify the distributions. 
We consider three basic classes: Gaussian
noise, astrophysical gravitational-wave signal, and a constant-frequency
detector artifact (''stationary line''), the two latter injected into the Gaussian
noise. 1D and 2D versions of a convolutional neural network classifier are
implemented, trained and tested on a broad range of signal frequencies. We
demonstrate that these implementations correctly classify the instances of data
at various signal-to-noise ratios and signal frequencies, while also showing
concept generalization i.e., satisfactory performance at previously unseen
frequencies. In addition we discuss the deficiencies, computational
requirements and possible applications of these implementations. 
\end{abstract}

\section{Introduction}
\label{sec:intro}
\subsection{Gravitational wave searches}

Gravitational waves (GWs) are distortions of the curvature of spacetime, propagating with the speed of light \cite{Einstein1916}. Direct experimental confirmation of their existence was recently provided by the LIGO and Virgo collaborations \cite{ALIGO2015,AdV2015} in the form of, till date, several binary black hole mergers \cite{2016PhRvL.116f1102A,2017PhRvL.118v1101A,2017PhRvL.119n1101A}, and one binary neutron star (NS) merger observations, the latter also electromagnetically bright \cite{2017PhRvL.119p1101A}; the first transient GW catalog \cite{2018arXiv181112907T} contains the summary of the LIGO and Virgo O1 and O2 runs. 

In addition to merging binary systems, among other promising sources of GWs are non-axisymmetric supernova explosions, as well as long-lived, almost-monochromatic GW emission by rotating, non-axisymmetric NS, sometimes called the ``GW pulsars''. 

In this article we will focus on the latter type of the signal. The departure from
axisymmetry in the mass distribution of a rotating NS can be caused
by dense-matter instabilities (e.g., phase transitions, r-modes), strong magnetic fields 
and/or elastic stresses in its interior (for a review see \cite{lasky2015,SieniawskaB2019}). 
The deformation and henceforth the amplitude of the GW signal depends on the largely unknown
dense-matter equation of state, surrounding and history of the NS, therefore the 
time-varying mass quadrupole required by the GW emission is not naturally guaranteed  
as in the case of binary system mergers. The LIGO and Virgo collaborations performed 
several searches for such signals, both targeted searches for NS sources of known spin 
frequency parameters and sky coordinates 
(pulsars, \cite{2017ApJ...839...12A,2017PhRvD..96l2006A} and references therein), 
as well as all-sky searches for {\it a priori} unknown sources with unknown parameters 
(\cite{2017PhRvD..96f2002A,2018PhRvD..97j2003A} and references therein). 

\subsection{All-sky searches for continuous GWs}

The {\it all-sky searches} for continuous GWs are `agnostic' in terms of GW frequency $f$, its 
time derivatives ({\it spindown} $\dot{f}$, sometimes $\ddot{f}$ and higher) and sky position of the source (e.g. $\delta$ and $\alpha$ in equatorial coordinates). The search consist in sweeping the parameter space to find the best-matching template by evaluating the signal-to-noise ratio (SNR). There are various algorithms (for a recent review of the methodology of continuous GW searches with the Advanced LIGO O1 and O2 data see \cite{Bejger2017,SieniawskaB2019}), but in the core they rely on performing Fourier transforms of the detectors' output time series.

Some currently-used continuous GW searches implement the
$\mathcal{F}$-statistic methodology \cite{jks1998}. In this work we will study
the output produced by of one of them, the all-sky time-domain $\mathcal{F}$-statistic
search \cite{AstoneBJPK2010} implementation, called the {\tt
TD-Fstat search}, \cite{polgraw-allsky} (see the documentation in
\cite{polgraw-allsky-docs}). This data analysis algorithm is based on matched
filtering; the best-matching template is selected by evaluating the SNR through maximisation of 
the likelihood function with respect to a set of above-mentioned frequency parameters $f$ and
$\dot{f}$, and sky coordinates $\delta$ and $\alpha$. By design, the $\mathcal{F}$-statistic 
is a reduced likelihood function \cite{jks1998,AstoneBJPK2010}. The remaining parameters characterizing the template  
- the GW polarization, amplitude and phase of the signal - do not enter the search directly, 
but are recovered after the signal is found. Recent examples of the use of 
the {\tt TD-Fstat search} include searches in the LIGO and Virgo data
\cite{VSR1TDFstat,2017PhRvD..96l2004A,2019arXiv190301901T}, as well as mock
data challenge \cite{2016PhRvD..94l4010W}. 

Assuming that the search does not take into account time derivatives higher
than $\dot{f}$, it is performed by evaluating the $\mathcal{F}$-statistic on a
pre-defined grid of $f$, $\dot{f}$, $\delta$ and $\alpha$ values in order to
cover the parameters space optimally and not to overlook the signal, whose true
values of $(f, \dot{f}, \delta, \alpha)$ may fall between the grid points.
The grid is optimal in the sense that for any possible signal there exists a grid point in the parameter space such that the expected value of the $\mathcal{F}$-statistic for the parameters of this grid point is greater than a certain value; for a detailed explanation see \cite{AstoneBJPK2010,PisarskiJ2015}.

The number of sky coordinates' grid points
as well as $\dot{f}$ grid points increases with frequency. Consequently the
volume of the parameter space (number of evaluations of the
$\mathcal{F}$-statistic) increases, see e.g., Fig. 4 in \cite{Poghosyan2015},
as well as the total number of resulting {\it candidate GW signals} (crossings
of the pre-defined SNR threshold) increases. For high frequencies, this type of
a search is particularly computationally demanding. 

The SNR threshold should preferably be as low as possible, because the continuous GWs are very weak - currently only upper limits for their strength are set \cite{2017PhRvD..96l2004A,VSR1TDFstat,2019arXiv190301901T,2017PhRvD..96f2002A,2018PhRvD..97j2003A}. A natural way to improve the SNR is to analyze long stretches of data since the SNR, denoted here by $\rho$, increases as a square root of the data length $T_0$: $\rho \propto \sqrt{T_0}$. In practice, coherent analysis of the many-months long observations (typical length of LIGO/Virgo scientific run is about one year) is computationally prohibitive. Depending on a method, the adopted coherence time ranges from minutes to days, then additional methods are used to combine the results incoherently. The {\tt TD-Fstat search} uses few-days long data segments for coherent analysis. In the second step of the pipeline the candidate signals obtained in the coherent analysis are checked for coincidences in a sequence of time segments to confirm the detection of GW \cite{VSR1TDFstat}. Here we explore an approach alternative to these studies, using a single data segment results to classify a distribution of candidate signals as potentially-interesting. In addition we note, that the coincidences step can be memory-demanding since the number of candidates can be very large, especially in the presence of spectral artifacts. The following work explores therefore an additional classification/flagging step for noise disturbances which can vastly reduce the number of signal candidates from a single time segment for further coincidences.

\subsection{The aim of this research}

The aim of this work is to classify the output of {\tt TD-Fstat search}, the multi-dimensional
distributions of candidate GW signals. Specifically, we study
the application of convolutional neural network (CNN) on the distribution of
candidate signals obtained by evaluating the {\tt TD-Fstat search} algorithm on
a pre-defined grid of parameters. The data contains either pure Gaussian noise, Gaussian
noise with injected astrophysical-like signals, or Gaussian noise with injected
purely monochromatic signals, simulating spectral artifacts local to the
detector (so-called stationary lines). 

\subsection{Previous works}

The CNN architecture \cite{Goodfellow2016} have already proved to be useful in the field of the GW physics, in particular in the domain of image processing. Razzano and Cuoco {\cite{Razzano2018}} have been using CNNs for classification of noise transients in the GW detectors. Beheshtipour add Papa \cite{2020PhRvD.101f4009B} have been studying the application of deep learning on the clustering of continuous gravitational wave candidates. George and Huerta{\cite{George2017}} have developed the {\it Deep Filtering} algorithm for the signal processing, based on a system of two deep CNNs, designed to detect and estimate parameters of compact binary coalescence signal in noisy time-series data streams. Dreissigacker et al. \cite{{2019arXiv190413291D}} have been using deep learning (DL) as a search method for the CWs from rotating neutron stars over broad range of frequencies, whereas Gebhard et al. \cite{2019arXiv190408693G} studied the general limitations of CNNs as a tool to search for merging black holes.

The last three papers discuss the DL as an alternative to matched filtering. However, it seems that the DL has too many limitations for the application in the classification of GW based on raw data from the interferometer (see discussion in \cite{2019arXiv190408693G}). For this reason we have decided to study a different application of DL. We consider DL as tool complementary to matched filtering, which allows to effectively classify large number of signal candidates obtained with the matched filter method. Instead of studying only binary classification, we have covered the multi-label classification assessing the case of artifacts resembling the CW signal. Finally our work compares two different types of convolutional neural networks implementations: one-dimensional (1D) and two-dimensional (2D).

\subsection{Structure of the article}

The article is organized as follows. In Sect.~\ref{sec:dl} we introduce the 
DL algorithms with particular emphasis on the convolutional neural
networks and their application in astrophysics. Section~\ref{sec:meth}
describes data processing we used to develop accurate model for the {\tt
TD-Fstat search} candidate classification. Section~\ref{sec:results} summarizes our
results which are further discussed. Summary
and a description of future plans are provided in Sect.~\ref{sec:conc}.

\section{Deep learning}
\label{sec:dl}

DL \cite{lecun_2015} has commenced a new area of machine learning, a field of computer science based on special algorithms that can learn from examples in order to solve problems and make predictions, without the need of being explicitly programmed \cite{Samuel:1959}. DL stands out as a highly scalable method that can process raw data without any manual feature engineering. By stacking multiple layers of artificial neurons (called neural networks) combined with learning algorithms based on back-propagation and stochastic gradient descent (\cite{Goodfellow2016} and references therein), it is possible to build advanced models able to capture complicated non-linear relationships in the data by composing hierarchical internal representations. The deeper the algorithm is, the more abstract concepts it can learn from the data, based on the outputs of the previous layers.

The DL is commonly used in commercial applications associated with computer vision {\cite{comp_vision}}, image processing {\cite{im_proc}}, speech recognition {\cite{speech_rec}} and natural language processing {\cite{nlp}}. What is more, it is also becoming more popular in science. The DL algorithms for image analysis and recognition have been successfully tested in many fields of astrophysics like galaxy classification {\cite{galaxy_class}} and asteroseismology {\cite{astero}}. Among many DL algorithms there is one that might especially be useful in the domain of the GW physics -- the CNNs. 

\subsection{Convolutional Neural Network}
\label{ssec:cnn_ss}

CNN is a deep, feed-forward artificial neural network (network that process the information only from the input to the output) whose structure is inspired by the studies of visual cortex in mammals, the part of the brain which specializes in processing visual information. The crucial element of CNNs is called a convolution layer. It detects local conjunctions of features from the input data and maps their appearances to a feature map. As a result the input data is split into parts, creating local receptive fields and compressed into feature maps. The size of the receptive field corresponds to the scale of the details to be looked in the data.

CNNs are faster than typical fully-connected \cite{dl_benchmark}, 
deep artificial neural networks because sharing weights significantly decreases the number of neurons required to analyze data. They are also less prone to the overfitting (the model learned the data {\it by heart} preventing the correct generalization). The {\it pooling layers} (subsampling layers)  coupled to the convolutional layers might be used to further reduce the computational cost. They constrain the size of the CNN and make it more resilient to the noise and translations which enhances their ability to handle new inputs.

\section{Method}
\label{sec:meth}

\subsection{Generation of data}
\label{ssec:gen_data}

To obtain a sufficiently large, labeled training set, we generate a set of {\tt TD-Fstat search} results (distributions of candidate signals) by injecting signals with known parameters. We define three different classes of signals resulting in the candidate signal distributions used subsequently in the classification: 1) a GW signal, modeled here by injecting an astrophysical-like signal that matches the $\mathcal{F}$-statistic filter, corresponding to spinning triaxial NS ellipsoid \cite{AstoneBJPK2010}, 2) an injected strictly-monochromatic signal, similar to realistic local artifacts of the detector (so-called stationary lines) \cite{2018PhRvD..97h2002C}, for which the $\mathcal{F}$-statistic is not an optimal filter, or 3) pure Gaussian noise, resembling `clean' noise output of the detector. These three classes are henceforth denoted by the {\bf cgw} (continuous gravitational wave), {\bf line} and {\bf noise} labels, respectively.   

\begin{figure}[htbp]
\centering
  \includegraphics[scale=0.7]{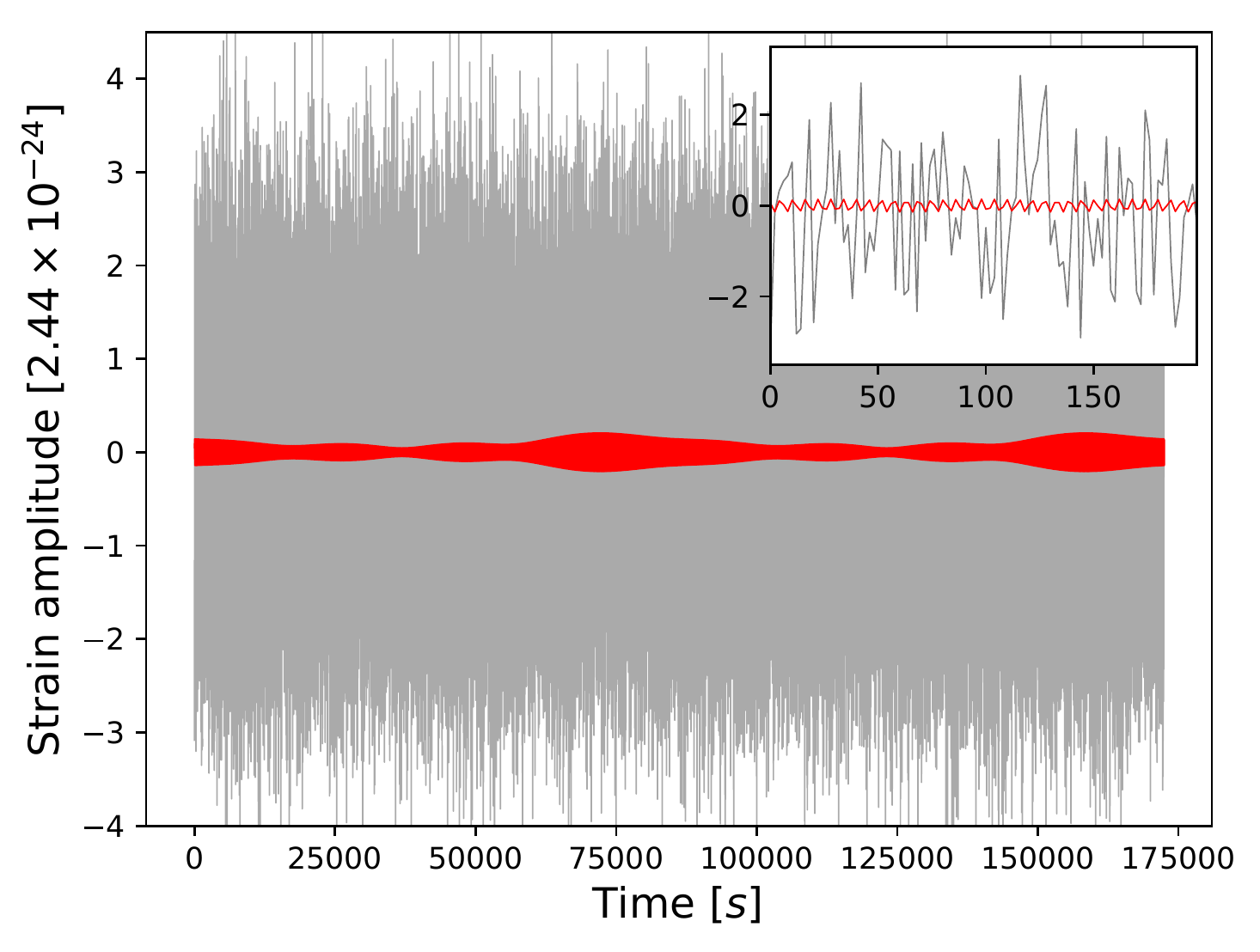}

  \caption{An example of a continuous GW time-domain data, input of {\tt TD-Fstat search}. Grey time series of $T_0$ = 2 sidereal days length mimics the downsampled, narrow-banded data produced from the raw interferometer data \cite{AstoneBJPK2010,VSR1TDFstat}. The data contains an almost-monochromatic astrophysical GW signal (red curve) of $\rho_{inj}=10$, and the following parameters (see also Table~\ref{tab:SearchParams} for the parameters of the search and the text for more details): frequency $f=2.16$ (in the units of the narrow band, between 0 and $\pi$), spindown $\dot{f}=-3.81\times 10^{-8}$ (in dimensionless units of the pipeline, corresponding to $\dot{f}_{astro}=-3.03\times 10^{-9}$ Hz/s; $\dot{f} = \pi\dot{f}_{astro}dt^2$ \cite{AstoneBJPK2010}), $\delta=0.474$ (range between $-\pi/2$ and $\pi/2$) and $\alpha=5.84$ (range between 0 and $2\pi$). The reference frequency of the narrow band equals 100 Hz. Visible modulation is the result of the daily movement of the detector with respect to the astrophysical source, as well as of their relative positions, reflecting the quadrupolar nature of the detector's antenna pattern; in the case of a stationary line local to the detector such modulation is absent.}

  \label{fig:raw_series} 
\end{figure}
 
To generate the candidate signals for the classification, the {\tt TD-Fstat search} uses narrow-banded time series data as an input. In this work we focus on stationary white Gaussian time series, into which we inject astrophysical-like signals, or monochromatic `lines' imitating local detector's disturbances. 
An example of such input data is presented in Fig.~\ref{fig:raw_series}. It simulates the raw data taken from the detector, downsampled from the original sampling frequency (16384 Hz in LIGO and 20000 Hz in Virgo) to 0.5 Hz, and be divided into narrow frequency bands. Because the frequency of an astrophysical almost-periodic GW signal is not expected to vary substantially (only by the presence of $\dot{f}$), we use a bandwidth of 0.25 Hz, as in recent astrophysical searches \cite{2017PhRvD..96l2004A,2019arXiv190301901T}. Each narrow frequency band is labeled by a reference frequency, related to the lower edge of the frequency band. Details of the input data are gathered in Table~\ref{tab:SearchParams}. Additional {\tt TD-Fstat search} input include the ephemeris of the detector (the position of the detector with respect to the Solar System Barycenter and the direction to the source of the signal, for each time of the input data), as well as the pre-defined grid parameter space of ($f$, $\dot{f}$, $\delta$. $\alpha$) values, on which the search ($\mathcal{F}$-statistic evaluations) is performed \cite{PisarskiJ2015}. 

In the signal-injection mode, the {\tt TD-Fstat search} implementation adds an artificial signal to the narrow-band time domain data at some specific $(f, \dot{f}, \delta, \alpha)_{inj}$, with an assumed signal-to-noise $\rho_{inj}$. For long-duration almost-monochromatic signals, which are subject of this study, $\rho_{inj}$ is proportional to the length of the time-domain segment $T_0$, the amplitude of the signal $h_0$ (GW 'strain') and inversely proportional to the amplitude spectral density of the data $S$, $\rho_{inj}=h_0\sqrt{T_0/S}$. The output SNR $\rho$ for a candidate signal corresponding to $(f, \dot{f}, \delta, \alpha)_{inj}$ is a result of the evaluation of the $\mathcal{F}$-statistic on the Gaussian-noise time series with injected signal. The value of $\rho$ at $(f, \dot{f}, \delta, \alpha)_{inj}$ is in general close, but different from $\rho_{inj}$ due to the random character of noise ($\rho$ is related to the value of $\mathcal{F}$-statistic as $\rho=\sqrt{2(\mathcal{F}-2)}$ (see \cite{2009agwd.book.....J} for detailed description). Furthermore it is calculated on a discrete grid. This is the principal reason why we do not study individual signal candidates and their parameters, but the resulting $\rho$ {\it distributions} in the $(f, \dot{f}, \delta, \alpha)$ parameter space (i.e. at the pre-defined grid of points), since the $\mathcal{F}$-statistic shape is complicated and has several local maxima, as shown e.g. in Fig. 1 of \cite{SieniawskaBK2019}. In the case of pure noise class no additional signal is added to the original Gaussian data, but the data is evaluated in pre-described range of $f, \dot{f}, \delta, \alpha$.

Subsequently to produce instances of the three classes for further classification, the code performs a search around the randomly-selected injection parameters $(f, \dot{f}, \delta, \alpha)_{inj}$, which in most cases fall in-between the grid points, in the range of a few nearest grid points ($\pm 5$ grid points, see Table~\ref{tab:SearchParams}). In case of {\bf cgw} all parameters are randomized, whereas for {\bf line} we take $\dot{f}\equiv 0$. To be consistent in terms of the input data e.g. number of candidate signals, in the case of a stationary line, we also select a random sky position and perform a search in a range similar to the {\bf cgw} case (this reflects a fact that spectral artifacts may also appear as clusters of candidate signal points in the sky). All the candidate signals crossing the pre-defined  $\mathcal{F}$-statistic threshold (corresponding to the SNR $\rho$ threshold) are recorded. 

\begin{figure}[htbp]
\centering
  \includegraphics[scale=0.4]{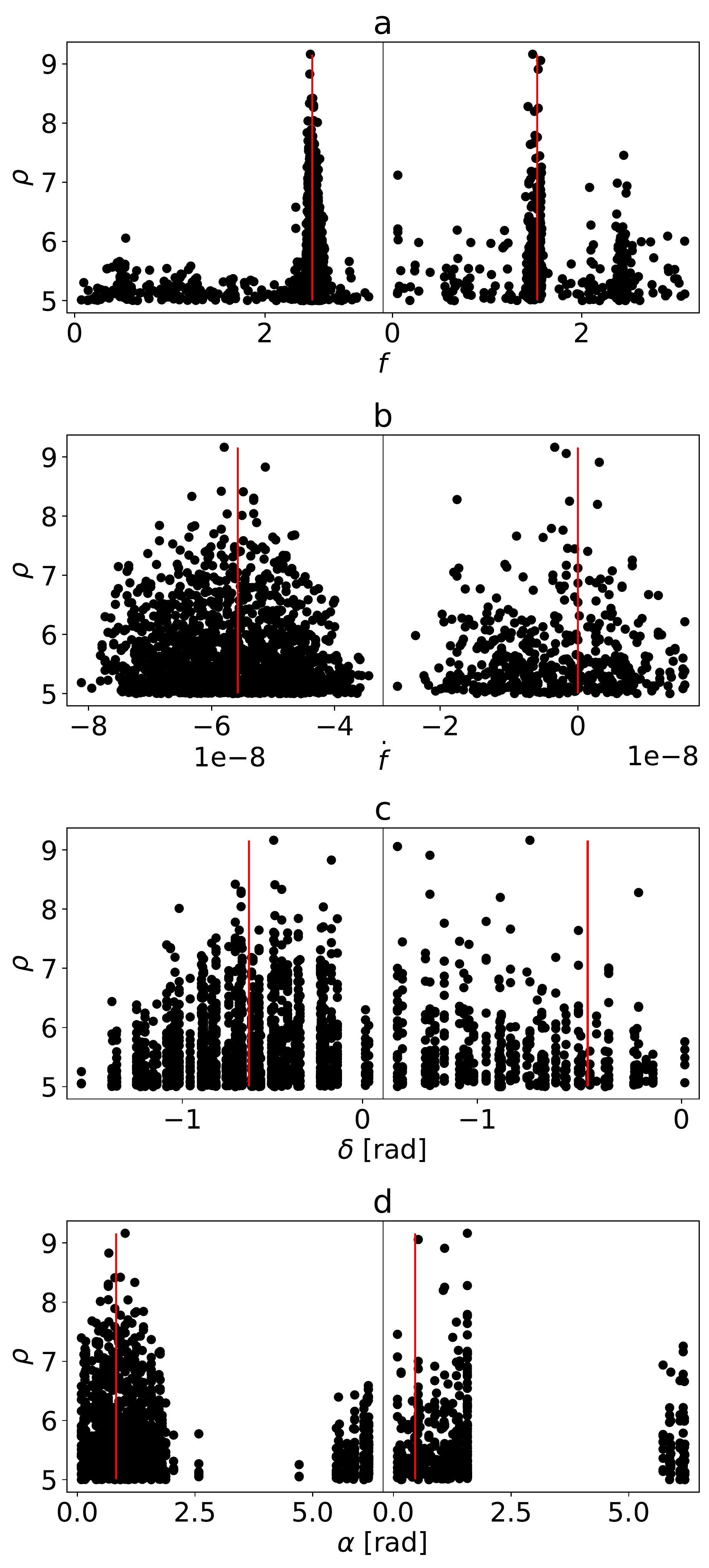}
  \caption{Exemplary {\tt TD-Fstat search} outputs for continuous GW signal and stationary line generated for the $\rho_{inj}=10$ and the reference band frequency $f=100$\,Hz and parameters of the injected signal from Fig.~\ref{fig:raw_series} ($f=2.16$, $\dot{f}=-3.81\times 10^{-8}$, $\delta=0.474$ and $\alpha=5.84$) marked by red vertical lines in the above plots. The left column corresponds to the {\bf cgw} whereas right column to the {\bf line}. The distributions represent the relation between recovered SNR $\rho$ ($\mathcal{F}$-statistic SNR) with respect to: the frequency -- $a$, the derivative of frequency -- $b$, the declination -- $c$ and the right ascension -- $d$. 
}    
  \label{fig:fstat_plots} 
\end{figure}

For each configuration of injected SNR $\rho_{inj}$ and reference frequency of the narrow frequency band, we have produced 2500 signals per class (292500 in total). 
For the {\bf cgw} class we assumed the simplest distribution over $\rho_{inj}$, i.e. a uniform distribution, as the actual SNR distribution of astrophysical signals is currently unknown. We apply the same `agnostic' procedure for the {\bf line} class; their real distribution is difficult to define without a detailed analysis of weak lines in the detector data (our methodology allows in principle to include such a realistic SNR distribution in the training set). To train the CNN, we put the lower limit of 8 on $\rho_{inj}$. Above this value, the peaks in the candidate signals $\rho$ distributions for the {\bf cgw} and {\bf line} classes are still visible on the $\rho(f, \dot{f}, \delta, \alpha)$ plots (see Fig. \ref{fig:fstat_plots} for the $\rho_{inj}=10$ case). For $\rho_{inj}<8$, the noise dominates the distributions hindering the satisfactory identification of signal classes. Nevertheless, in the testing stage of the algorithm we extend the range of $\rho_{inj}$ down to 4.

To summarize, each instance of the training classes is a result of the following input parameters: $(f, \dot{f}, \delta, \alpha)_{inj}$ and $\rho_{inj}$, and consist of a resulting distribution of the candidate signals: values of the SNR $\rho$ evaluations of the {\tt TD-Fstat search} at the grid points of the frequency $f$ (in fiducial units of the narrow-band, from 0 to $\pi$), spindown $\dot{f}$ (in Hz/s), and two angles describing its sky position in equatorial coordinates, right ascension $\alpha$ (values from 0 to $2\pi$) and declination $\delta$ (values from $-\pi/2$ to $\pi/2$); see Fig. {\ref{fig:fstat_plots}} for an exemplary output distribution of the candidate signals.

\begin{table} [htbp]
\centering
\begin{tabular}{ l | c }
  Detector & LIGO Hanford \\ \hline
  Reference band frequency & 50, 100, 200, 300, 500, 1000 Hz \\ & (20, 250, 400, 700, 900 Hz for tests) \\ \hline  
  Segment length $T_0$ & 2 days \\ \hline 
  Bandwidth & 0.25 Hz \\ \hline 
  Sampling time $dt$ & 2 s \\ \hline 
  Grid range & ${\pm}5$ points \\ \hline
  $\mathcal{F}$-statistic (SNR) threshold & 14.5 (corresponding to $\rho=5$)\\ \hline 
  Injected signal-to-noise $\rho_{inj}$ & from 8 to 20 \\ & (from 4 to 20 for tests)
\end{tabular}
  \caption{Parameters of the input to the {\tt TD-Fstat search} code (see e.g., \cite{VSR1TDFstat}). Time series consist initially of random instances of white Gaussian noise, to which {\bf cgw}s or {\bf line}s were added. Segment length $T_0$ equal to 2 sidereal days with 2 s sampling time results in 86164 data points. The $\mathcal{F}$-statistic (SNR) threshold is applied in order to select signal candidates above certain SNR ratio, to exclude those that are most likely a result of random noise fluctuations. 
  }
\label{tab:SearchParams} 
\end{table}

The CNN required the input matrix of the fixed size. However, the number of points on distributions shown in Fig.~\ref{fig:fstat_plots} may vary for each simulation. Depending on the frequency (see Table~\ref{tab:SearchParams}) it may increase a few times. To address this issue, we transformed point-based distributions into two different representations: set of four 2D images (four distributions) and set of five 1D vectors (five $\mathcal{F}$-statistic parameters).

\begin{figure}[htbp]
\centering
  \includegraphics[scale=0.5]{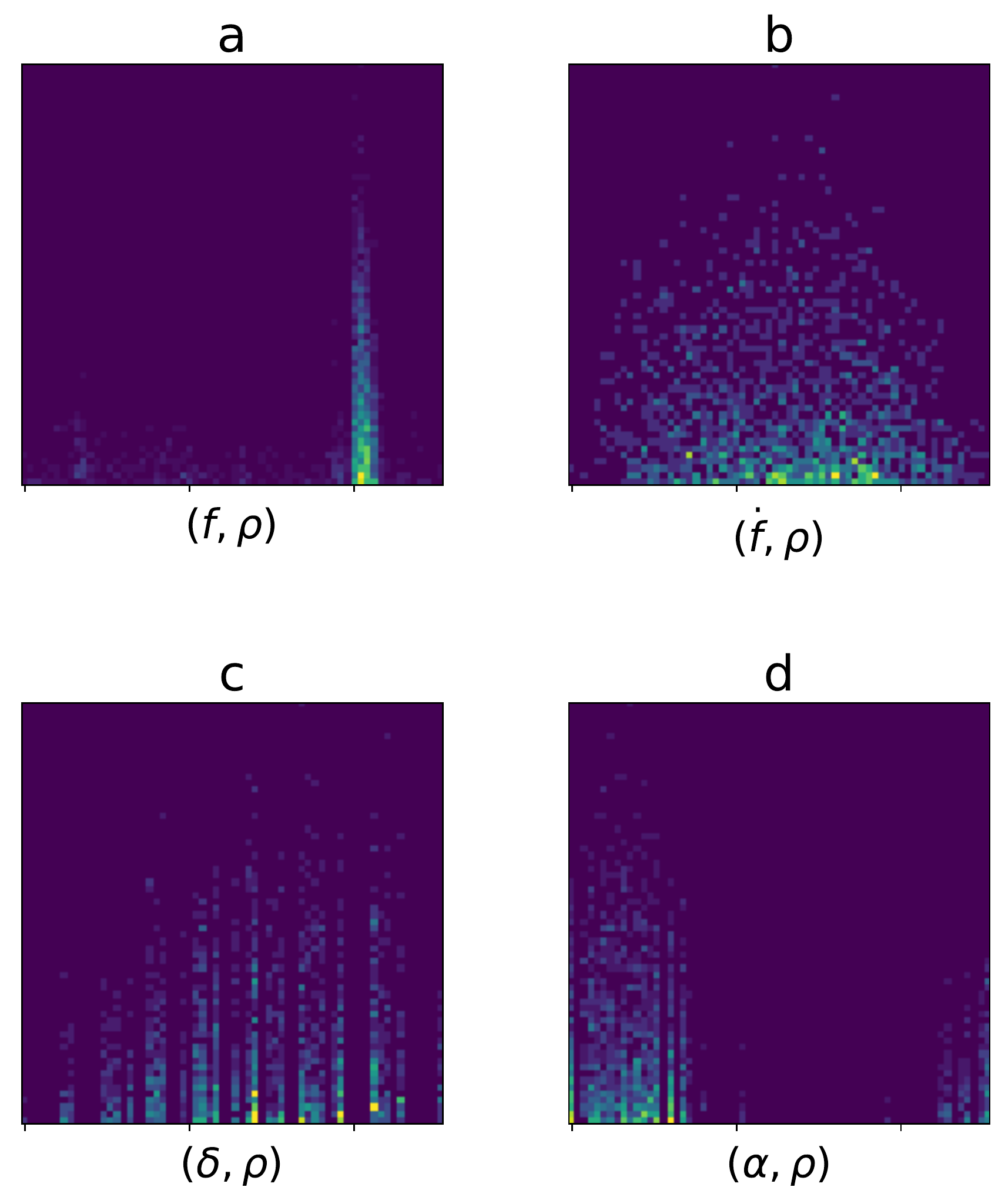} 
  \caption{The 2D representation of {\tt TD-Fstat search} outputs used as an input to the 2D CNN. Images presented here have same size equal to 64x64 pixels. They correspond to the distributions shown on the left column of Fig. \ref{fig:fstat_plots}: $a$ -- (frequency, $\rho$), $b$ -- (spindown, $\rho$), $c$ -- (declination, $\rho$), $d$ -- (right ascension, $\rho$). The colours correspond to the density of the distribution -- the brighter it is, the more points contributed to the pixel.}
  \label{fig:fstat_images} 
\end{figure}

The image-based representation was created via conversion to the two-dimensional histogram (see Fig. \ref{fig:fstat_images}) of the corresponding point-based distributions. Their sizes are $64 \times 64$ pixels. We chose this value empirically; smaller images lost some information after the transformation, whereas bigger images led to the significantly extended training time of the CNN we used.

\begin{figure}[htbp]
\centering
  \includegraphics[scale=0.4]{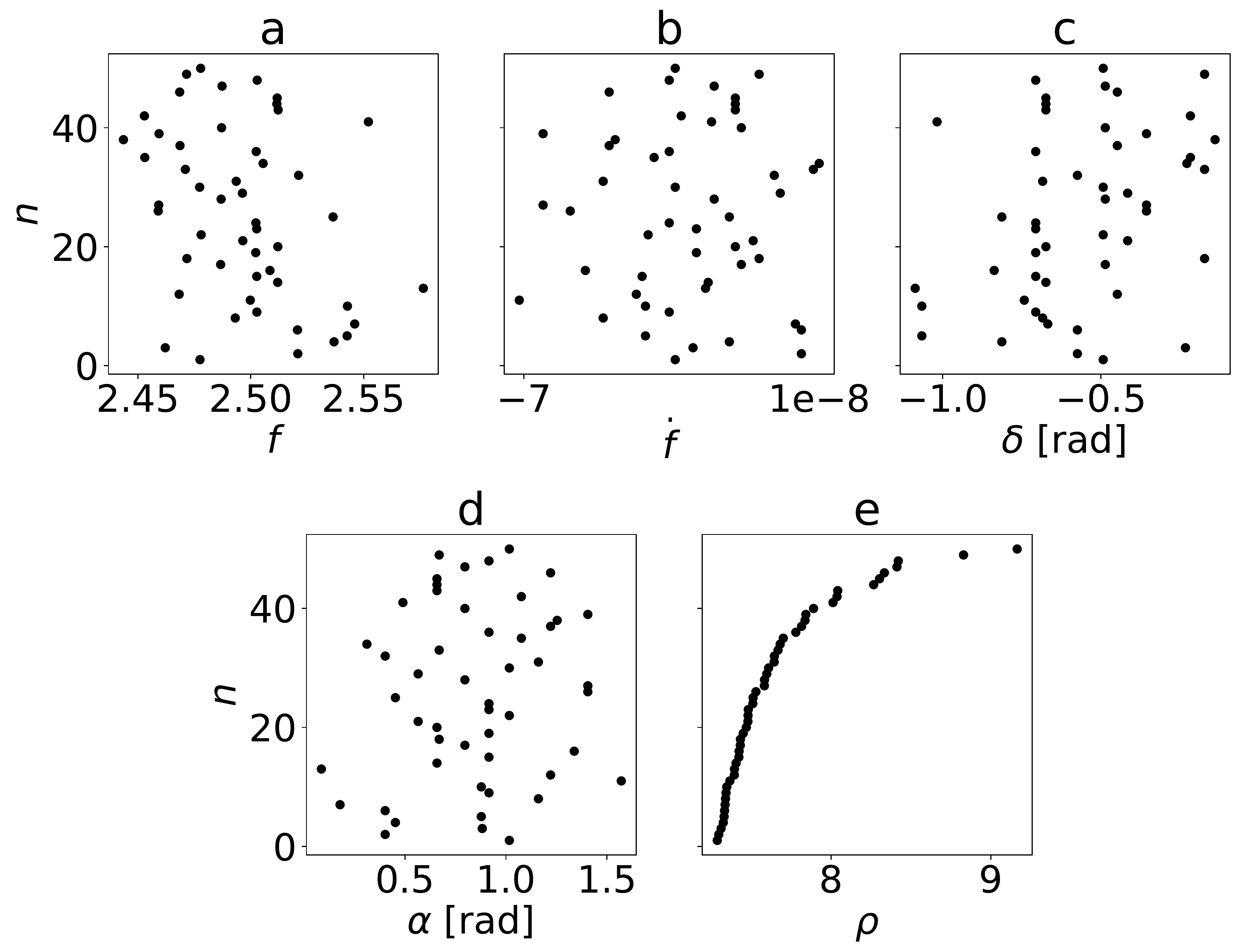} 
  \caption{The 1D representation of {\tt TD-Fstat search} outputs used as an input to the 1D CNN. The outputs are limited to the 50 maximum values of $\rho$ (plots presented here correspond to the distributions shown on the left column of Fig. \ref{fig:fstat_plots}$a$): $a$ -- frequency, $b$ -- spindown, $c$ -- declination, $d$ -- right ascension and $e$ -- SNR $\rho$. The vector of $\rho$ was sorted since it allowed to reach higher accuracy during training.}
  \label{fig:fstat_vectors} 
\end{figure}

The vector-based representation was created through the selection of the $50$ greatest values of the $\rho$ distribution and their corresponding values from the other parameters ($f$, $\dot{f}$, $\delta$ and $\alpha$). The length of the vector was chosen empirically. The main limitation was related to the density of the point-like distributions which changed proportionally to the frequency. For the 50 Hz signal candidates, the noise class had sparse distributions of slightly more than 50 points. Furthermore, the vectors were sorted with respect to the $\rho$ values, see Fig.~\ref{fig:fstat_vectors}: this step allowed to reach slightly higher values of classification accuracy. 

The created datasets were then split into three separate subsets: training set (60\% of signals from the total dataset), validation set (20\% of signals from the total dataset) and the testing set (20\% of signals from the total dataset). Validation part was used during training to monitor the performance of the network (whether it overfits). Testing data was used after training to check how the CNN performs with unknown samples.

\subsection{Neural network architecture}
\label{ssec:network_arch}

\begin{figure}[htbp]
\centering
  \includegraphics[scale=1.0]{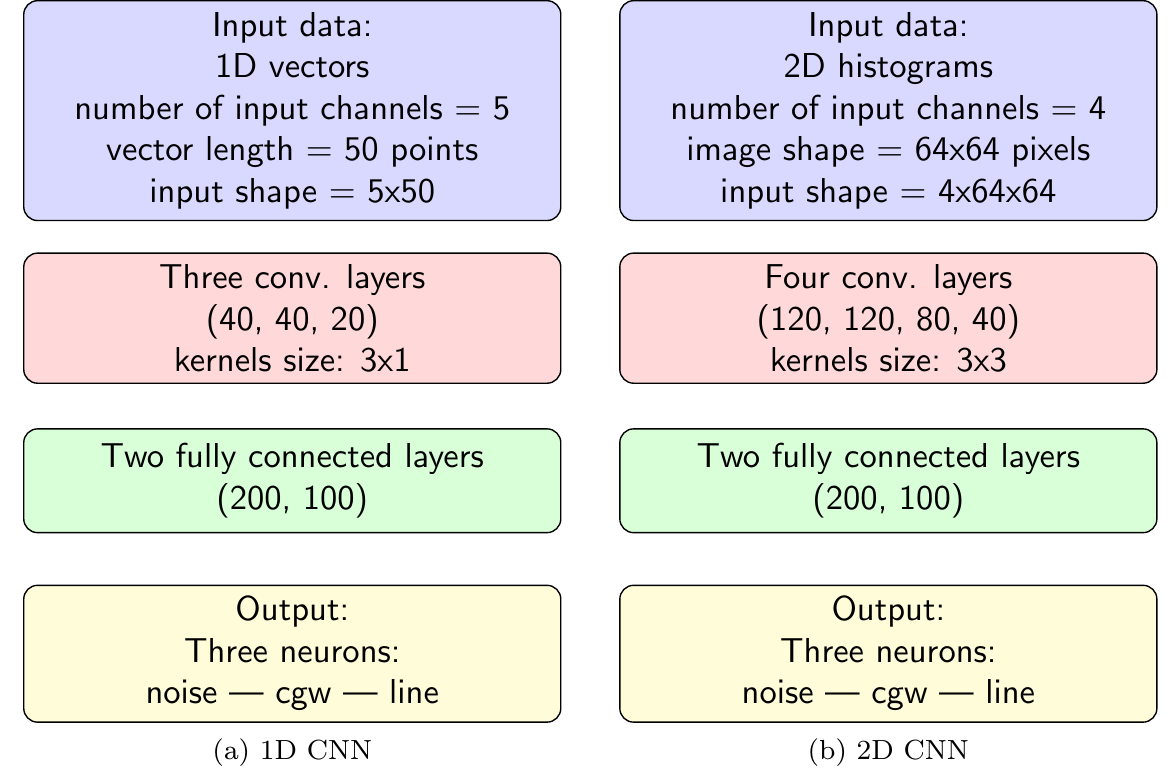}
  \caption{Diagrams show the networks' layer structure and architecture.}    
  \label{fig:arch} 
\end{figure}

The generated datasets required two different implementations of the CNN. Overall we tested more than 50 architectures ranging $2-6$ convolutional layers and $1-4$ fully connected layers for both models. Final layouts are shown in Figs. {\ref{fig:arch}$a$} and {\ref{fig:arch}$b$}. The architectures which were finally chosen are based on the compromise between the model accuracy and the training time. Models larger than those specified in Fig. {\ref{fig:arch}$a$} and {\ref{fig:arch}$b$} achieved similar performance, but at the cost of significantly longer training time.

In case of 1D CNN, the classifier containing three convolutional layers and two fully connected layers yielded the highest accuracy (more than 94\% for the whole validation/test datasets). Whereas the 2D CNN required four convolutional layers and two fully connected layers to reach the highest accuracy (85\% over the whole validation/test datasets). The models were trained for 150 epochs which took 1 hour for 1D CNN and 15 hours for 2D CNN (on the same machine equipped with the Tesla K40 NVidia GPU).

To avoid overfitting we included dropout \cite{dropout} in the architecture of both models. The final set of hyperparameters used for the training was the following for both implementations (definitions of all parameters specified here can be found in \cite{Goodfellow2016}): {\tt ReLU} as the activation function for hidden layers, {\tt softmax} as the activation function for output later, {\tt cross-entropy} loss function, {\tt ADAM} optimizer \cite{ADAM2014}, batch size of 128, and 0.001 learning rate (see Fig. \ref{fig:arch}$a$ and \ref{fig:arch}$b$ for other details). The total number of parameters used in our models were the following: 52503 for the 1D CNN, and 398083 for the 2D CNN.

The CNN architectures were implemented using the Python Keras library \cite{chollet2015keras} on top of the Tensorflow library \cite{tensorflow}, with support for the GPU. We developed the model on NVidia Quadro P6000\footnote{Benefiting from the donation via the NVidia GPU seeding grant.} 
and performed the production runs on the Cyfronet Prometheus cluster\footnote{Prometheus, Academic Computer Centre CYFRONET AGH, Kraków, Poland} equipped with Tesla K40 GPUs, running CUDA 10.0 \cite{cuda} and the cuDNN 7.3.0 \cite{cuDNN}.

\section{Results and discussion}
\label{sec:results}

\begin{figure}[htbp]
\centering
  \includegraphics[scale=0.5]{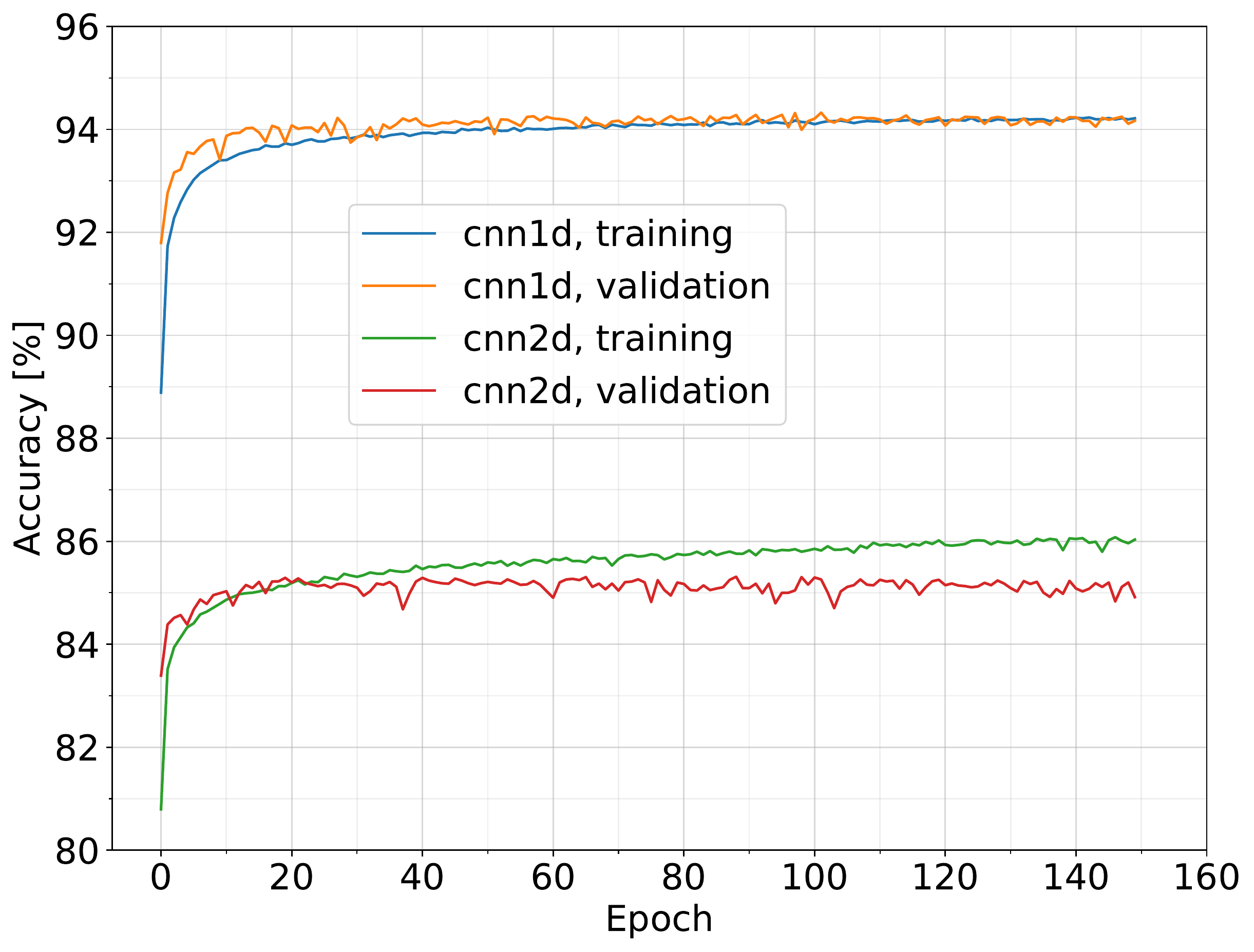}
  \caption{
  The evolution of the accuracy as a function of the training epoch for the three-label classification for the 1D CNN (upper curves) and 2D CNN (lower curves). Both models reached the maximum accuracy after 40 epochs (based on the results on the validation set). We prolonged training to 150 epochs to investigate the onset of overfitting. The 1D CNN was still properly learning (although without increase of validation accuracy), whereas the 2D CNN showed overfitting - validation accuracy (red curve) maintained at the constant level when training accuracy (green curve) was increasing.
}    
  \label{fig:training_both_cnn} 
\end{figure}

Both CNNs described in Section \ref{ssec:network_arch}, Fig. \ref{fig:arch}$a$ and Fig. \ref{fig:arch}$b$ were trained on the generated datasets. During the training model implementing 1D architecture was able to correctly classify 94\% of all candidate signals, whereas the model implementing 2D architecture reached 85\% of accuracy (see the comparison between learning curves in Fig. \ref{fig:training_both_cnn}). Accuracy is defined as the fraction of correctly predicted instances of data to total number of signal candidates. Since the very first epoch, the first model showed better ability to generalize candidate signals over large range of frequencies and values of injected SNR $\rho_{inj}$.

\begin{figure}[htbp]
    \centering
    \includegraphics[scale=0.5]{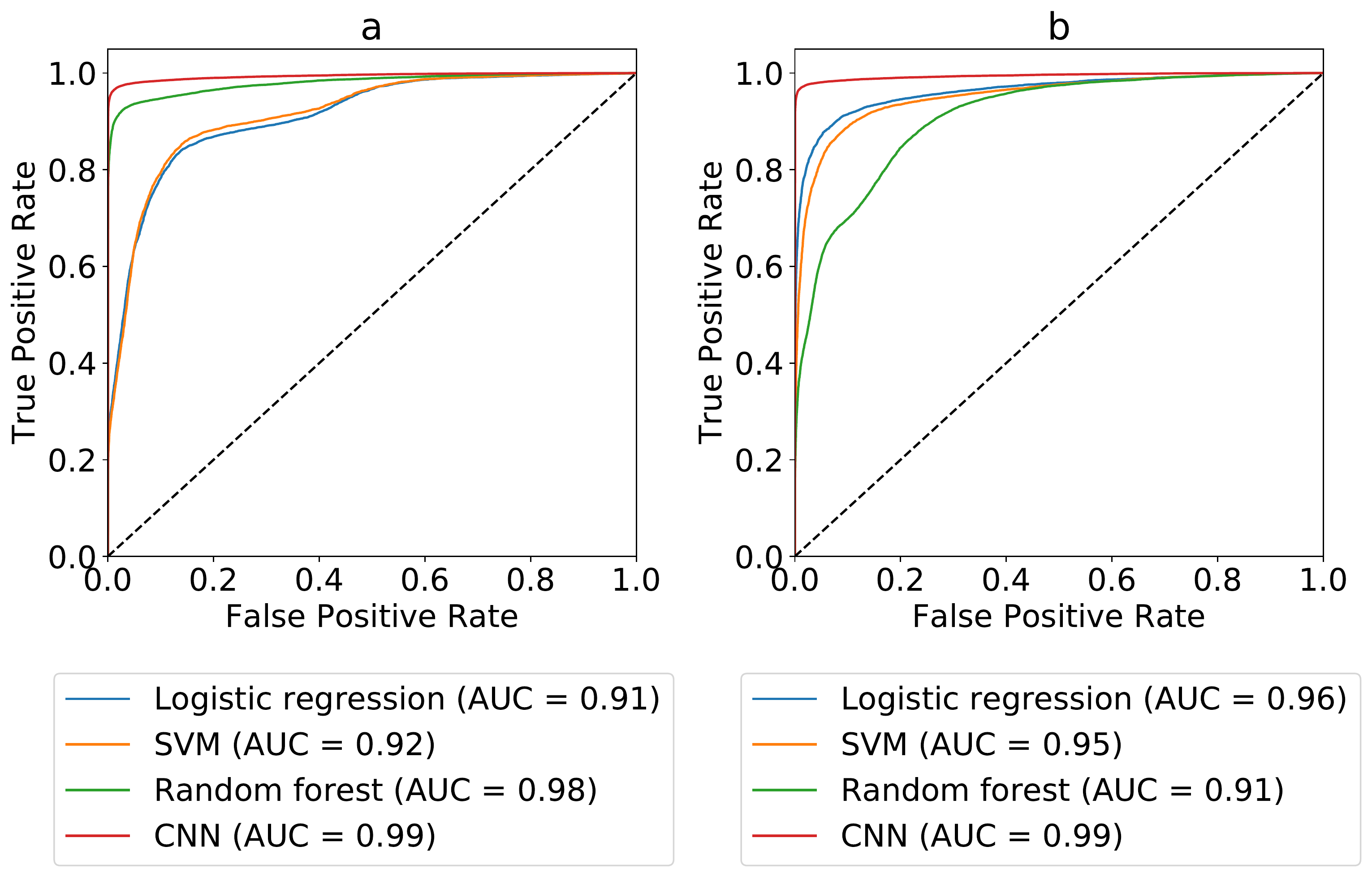}
    \caption{The Receiver-Operating-Characteristic (ROC) curves for compared ML algorithms trained on 1D data representation ($a$) and 2D data representation ($b$). The presented results are for the binary classification problem in which $Positive$ stands for {\bf cgw} whereas $Negative$ corresponds to combined {\bf line} and {\bf noise} class. Black dashed line in the middle correspond to random guessing (AUC stands for the Area Under Curve).}
    \label{fig:roc}
\end{figure}

To justify the choice of CNN as an algorithm suitable for the classification of signal candidates, we made a comparison test with different ML methods such as logistic regression, support vector machine (SVM) and random forest. For the test we modified the multi-label classification problem into binary case to create Receiver-Operating-Characteristic (ROC) curves. Classes of {\bf line} and {\bf noise} were combined into a single non-astrophysical class. The results of the comparison are shown on Figs.~\ref{fig:roc}$a$ and \ref{fig:roc}$b$. The results shown on the left figure are corresponding to models trained and tested on 1D data representation, whereas the results shown on the right plot refer to 2D data representation. In both cases the CNNs outperformed other ML models. To further underline the differences, Tab. \ref{tab:roc_pdet} shows the detection probability (True Positive Rate, TPR) at a $1\%$ of false alarm rate (False Positive Rate or FPR).

CNNs achieved similar level of detection probability significantly outperforming the other algorithms. In case of binary classification or detection of {\bf cgw} 2D CNN seemed to be slightly better even with much lower accuracy as shown on Fig. \ref{fig:training_both_cnn}. However the aim of our work was not only to classify GWs, but also to investigate it's usefulness on the detection of the stationary line artifacts. The data collected by the GW detectors is noise dominated and polluted by spectral artifacts in various frequency bands, which significantly impact the overall quality of data. Since the CNNs may potentially help in classification of lines to remove them from the science data, the analysis with respect to the multi-label problem is beneficial.

\begin{table}[h!]
\begin{center}
    \begin{tabular}{|c|c|c|c|c|}
      \hline
      \multicolumn{5}{|c|}{Detection probability of {\bf cgw} at 1\% false alarm rate}\\
      \hline
      \hline
      \textbf{} & \textit{\, Logistic regression \, } & \textit{SVM} & \textit{\, Random Forest \, } & \textit{CNN}\\
      \hline
      1D data & $33.8$ & $31.9$ & $89.2$ & $96.3$ \\
      \hline
      2D data & $72.8$ & $57.6$ & $38.4$ & $96.8$ \\
      \hline
    \end{tabular}
    \caption{The summary of detection probabilities for {\bf cgw} at 1\% false alarm rate for compared ML algorithms trained and tested on 1D and 2D data representations. }
    \label{tab:roc_pdet}
  \end{center}
\end{table}

\begin{figure}[htbp]
\centering
  \includegraphics[scale=0.5]{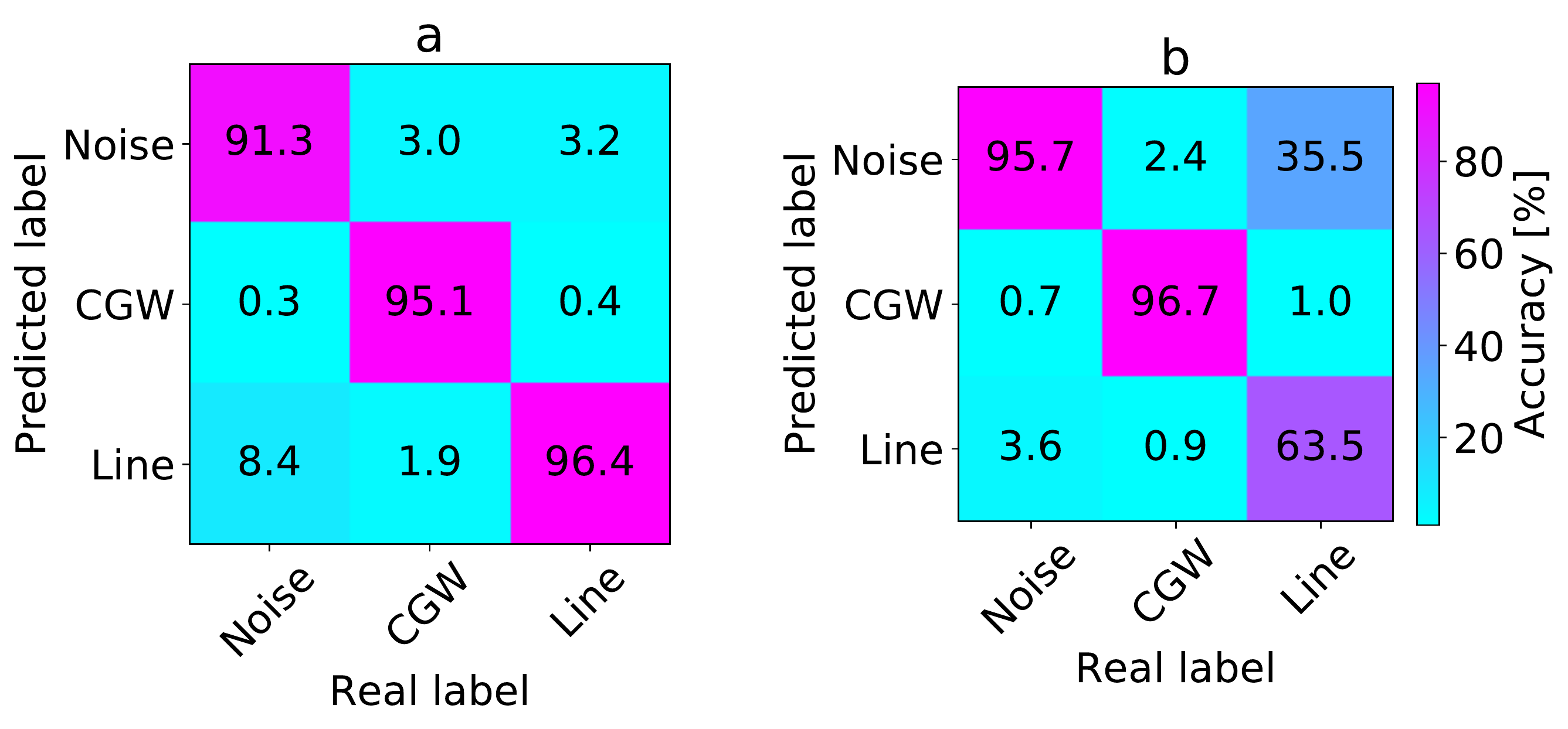}
  \caption{
  Confusion matrix for the three-label classification evaluated on the test set for the 1D CNN -- $a$ and 2D CNN -- $b$ after the training. Although the {\bf cgw} and the {\bf noise} were classified on the similar level, the {\bf line} caused significant problem for the 2D model. Majority of {\bf line} instances resembled {\bf noise} class in the image representation.
}    
  \label{fig:cm_test} 
\end{figure}

To decide which CNN architecture was more suitable to the multi-classification, our models were tested against unknown before samples (test dataset), after the training. The results were shown in Fig. \ref{fig:cm_test} in the form of confusion matrix. Both models were able to correctly classify majority of {\bf cgw} (95.1\% for the 1D model and 96.7\% for 2D model) as well as the {\bf noise} (91.3\% and 95.7\% respectively). However the difference in the classification of the {\bf line} was significant. 1D CNN was able to correctly classify 96.4\% of line candidates whereas 2D CNN only 63.5\%. 
Although the 2D model seemed to be more suited for binary classification task (detection of GW signal from the noise), the 1D CNN outperformed 2D version in the multi-label classification.

Knowing the general capabilities of designed CNNs, we performed additional tests trying to understand the response of our models against signal candidates of the specific parametrization. We generated additional datasets for particular values of the SNR $\rho_{inj}$ and the frequency (see Table~\ref{tab:SearchParams}). We expanded the $\rho$ range down to value of 4 which corresponds to the $\mathcal{F}$-statistic threshold for the signal candidate. This step allowed us to test the response of the CNN against unknown during training very weak signals that seemed to be indistinguishable from the noise.

\begin{figure}[htbp]
\centering
  \includegraphics[scale=0.5]{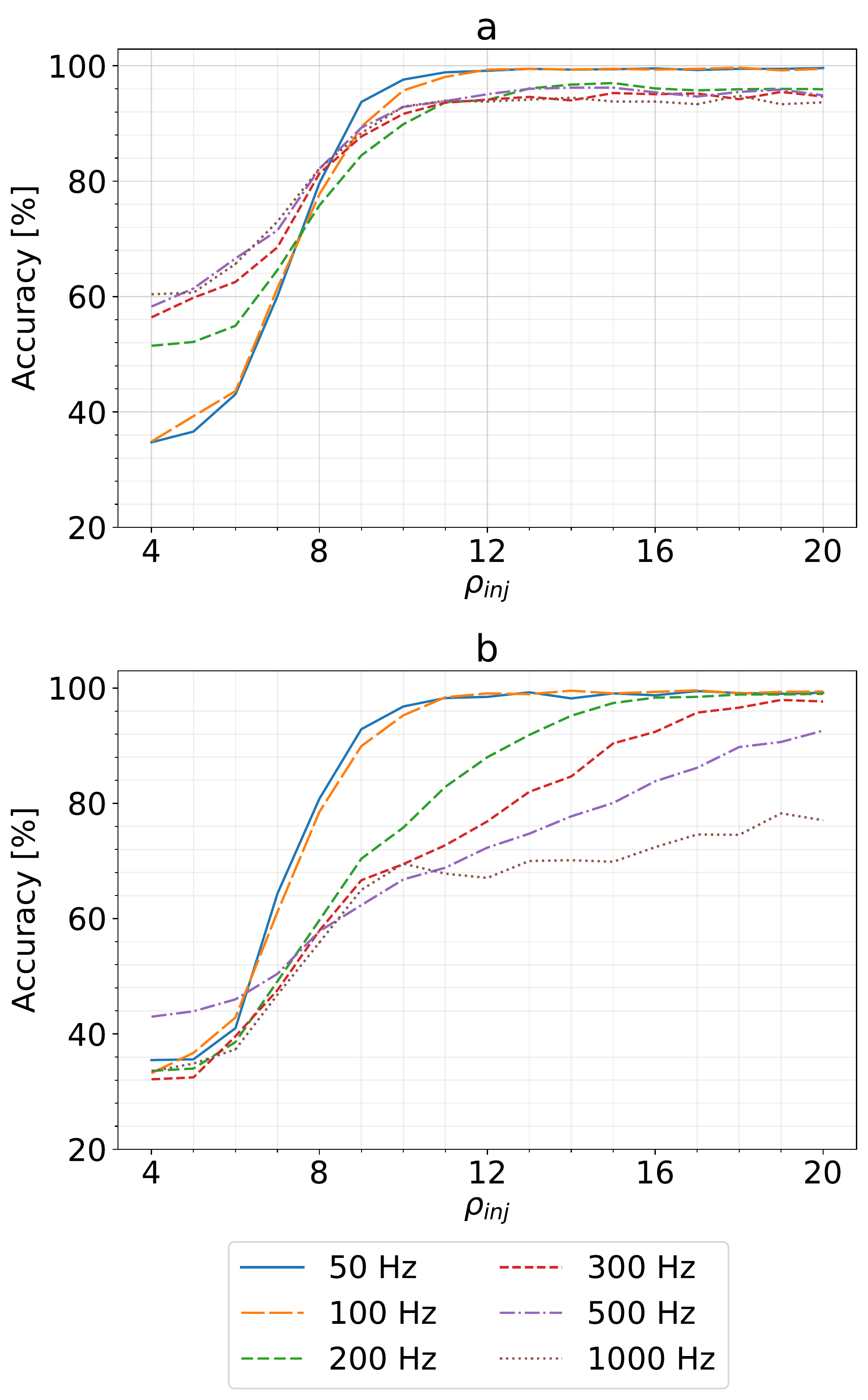}
  \caption{The evolution of accuracy as the function of the injected SNR $\rho_{inj}$ for 1D CNN ($a$) and 2D CNN ($b$). The first model achieved maximum level of accuracy for $\rho_{inj}=10-12$ and maintained its value for whole range of frequencies. The 2D version varied significantly in relation to the frequency with the maximum accuracy being gradually shifted toward larger values of $\rho_{inj}$. Characteristic shift in the accuracy (upper plot) between the lower frequencies (50 and 100 Hz) and the rest was associated with the density of signal candidates distributions. {\bf cgw} and {\bf line} instances were easier to separate from {\bf noise} since their distributions of parameters had very sparse character (see Fig \ref{fig:fstat_plots} for comparison) - the {\bf noise} signal candidates were not grouping around fluctuations in the frequency domain (the background of Fig.~\ref{fig:fstat_plots}$a$) allowing easier classification than for higher frequencies.
}    
  \label{fig:snr_acc_comparison} 
\end{figure}

The results were presented in Figs.~\ref{fig:snr_acc_comparison}$a$ and \ref{fig:snr_acc_comparison}$b$ (for 1D and 2D CNNs, respectively). The 1D model presented significantly more stable behaviour toward the candidates over whole range of considered frequencies. It also maintained nearly stable accuracy for the data with the injected SNR $\rho_{inj}{\geq}10$ (reaching the value of more than 90\% for all of them). Interestingly, candidates with $\rho_{inj} < 8$ were correctly classified in $60-70$\% of samples for frequency $\geq$ 200 Hz. This was relatively high value, taking into consideration their noise-like pattern (for {\bf cgw} and {\bf line} instances). This pattern had the biggest influence on the classification of the signal candidates generated for frequencies: 50, 100 Hz and the $\rho_{inj} < 8$. The small number of points contributing to the peak (see Fig.~\ref{fig:fstat_plots}$a$ for comparison) with respect to the background noise, made these candidates hardly distinguishable from the {\bf noise} class.

On the other hand, 2D CNN varied significantly in relation to the frequency. It reached the highest accuracy for the 100 Hz (99\% for the $\rho_{inj}>10$). For the other frequencies, the maximum accuracy was gradually shifted toward increasing $\rho_{inj}$. Interestingly, the accuracy for the 50 Hz reached the maximum for the $\rho_{inj}=10$; then it gradually decreased. The 2D CNN seemed to outperform 1D model only for the narrow band of the frequency. Nevertheless, the general performance of this implementation was much worse. 

Since the 1D CNN proved to be more accurate over broad range of frequencies, we chose it as a more useful model in the classification of the $\mathcal{F}$-statistic signal candidates. Below we present the results of additional tests we performed to better understand its usability.

\begin{figure}[htbp]
\centering
  \includegraphics[scale=0.5]{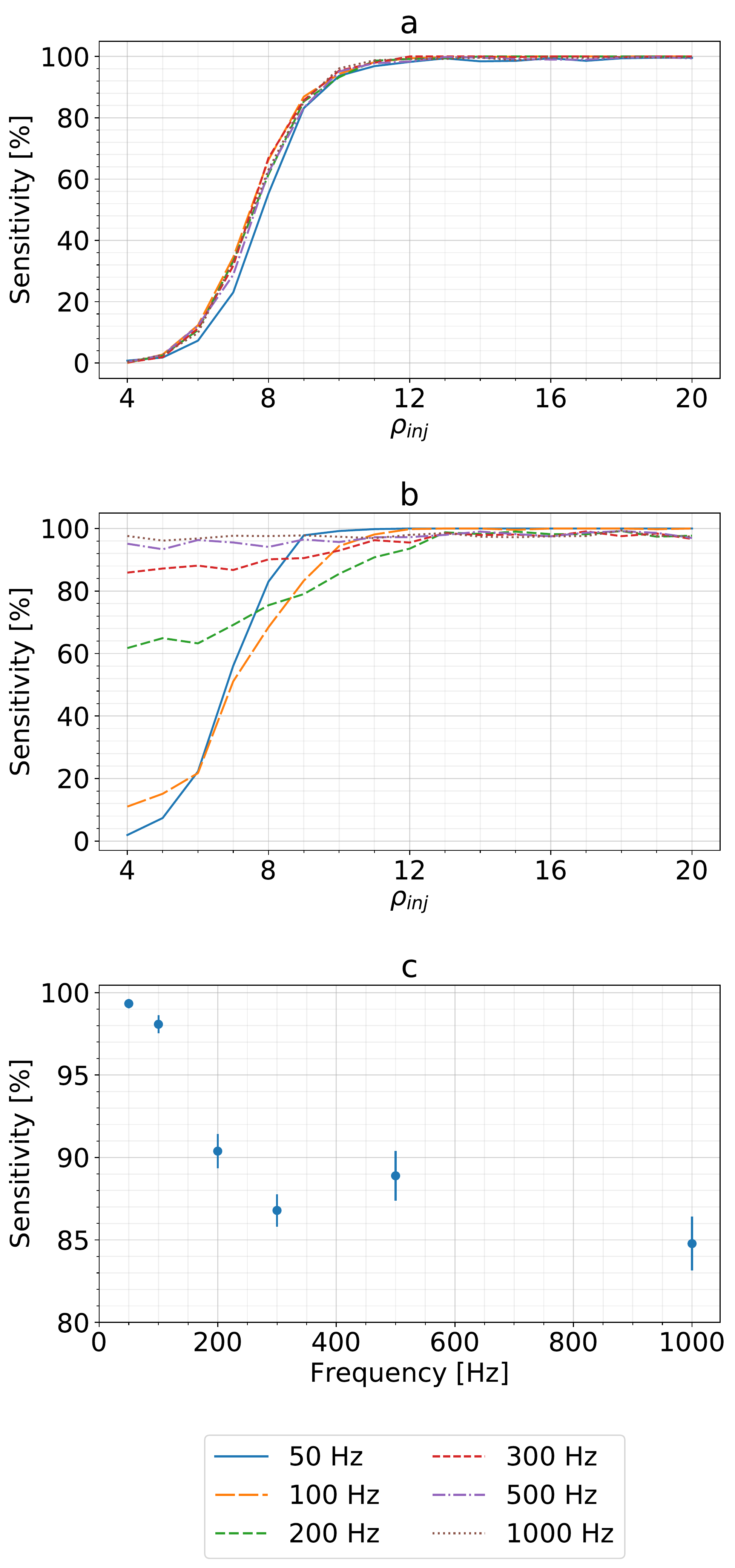}
  \caption{The evolution of sensitivity as a function of SNR $\rho_{inj}$ of the 1D CNN for the three types of signal candidates: {\bf cgw} -- $a$, {\bf line} -- $b$, and {\bf noise} -- panel $c$. The last panel shows average values for frequencies, because the {\bf noise} classification sensitivity is not a function of the injected SNR $\rho_{inj}$, and stays approximately constant for each narrow-band frequency value.
}    
  \label{fig:recall} 
\end{figure}

To tests the model response toward particular signal candidate, we computed sensitivity (in ML literature also referred as the recall) defined as the fraction of relevant instances among the retrieved instances. Figure \ref{fig:recall} presents the results. 
Classification of the {\bf cgw} was directly proportional to the $\rho_{inj}$ up to value of 11-12, then the sensitivity saturated around 95\%-99\% depending on the frequency. For $\rho_{inj}$ approaching 4, sensitivity decreased to 0\%. This result was expected since the injected signal at this level is buried so deeply in the noise that is indistinguishable. Furthermore by comparing Fig. \ref{fig:recall} $a$ with Fig. \ref{fig:snr_acc_comparison} $a$, we deduced that the classification of {\bf cgw} had the biggest influence on the total performance of the CNN. 

The sensitivity of the {\bf line} for higher frequencies (more than 300 Hz) maintained at relatively constant level of more than 95\% even for the smallest $\rho_{inj}$. Decrease in sensitivity for lower frequencies was associated with the density of the signal candidates distribution. The outputs of {\tt TD-Fstat} had the more sparse character, the lower frequency was. Chosen 50 points for the input data
were taken not only from the peak but also from the background noise (see top plots from Fig. \ref{fig:fstat_plots}). With decreasing $\rho_{inj}$ background points started to dominate and the candidates seemed to resemble {\bf noise} class. This leads to misclassification of nearly all {\bf line} samples for 50 Hz data.

In case of the {\bf noise}, sensitivity was inversely proportional to the frequency. Again this was associated with the density of the signal candidates distributions. For higher frequencies more points contributed to local fluctuations. As a result the 50 points chosen for the input data, instead of having random character, resembled different types of candidates.

\begin{figure}[htbp]
\centering
  \includegraphics[scale=0.4]{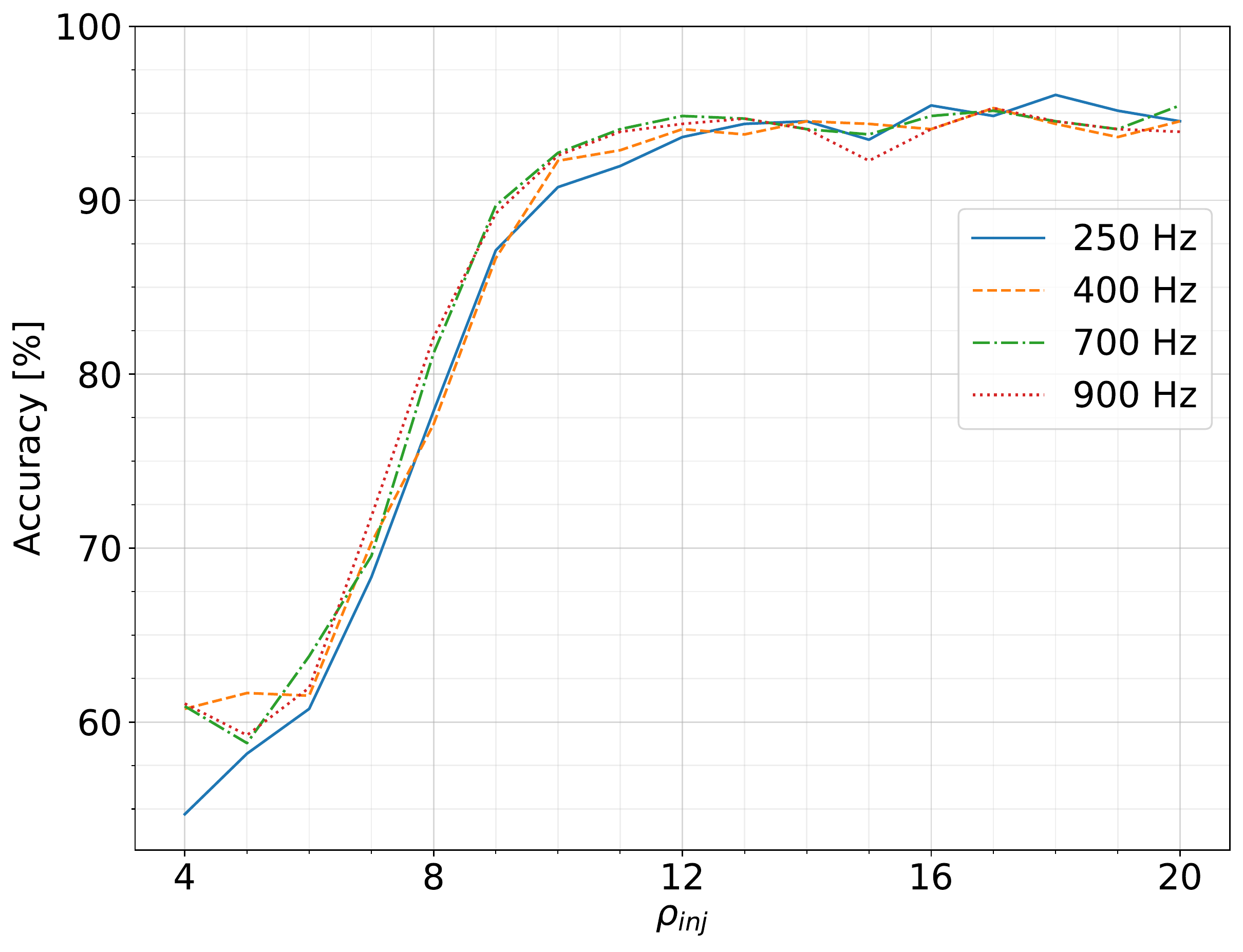}
  \caption{The evolution of accuracy as the function of the injected SNR $\rho_{inj}$ for 1D CNN for the signal candidates generated with frequencies different from those used for the training.
}    
  \label{fig:snr_acc_newfreq_cnn1d} 
\end{figure}

We additionally performed tests on the signal candidates generated for different frequencies than specified in the Table~\ref{tab:SearchParams}. We chose five new frequencies to test the model on: 20, 250, 400, 700, 900 Hz. The results were presented in Fig. \ref{fig:snr_acc_newfreq_cnn1d}. The 20 Hz case is missing since the number of available points (from initial distributions) to create set of five 1D vectors was much smaller than the chosen length (some distributions for the {\bf noise} class contained less than 10 points). Nevertheless, the CNN for the other frequencies reached similar accuracies as those presented in Fig. \ref{fig:snr_acc_comparison}$a$. This result proved the generalization ability of the 1D CNN toward unknown frequencies. However the limitation of the model was the minimum number of candidate signals available to create input data. Since this number was proportional to the number of grid points (frequency) of the searched signal, our CNN was not suited to search for candidates below 50 Hz.

Although it is not immediately apparent from the 1D and 2D instances of the distributions of candidate signals, the $\mathcal{F}$-statistic values in the sky points contain non-negligible information about the signal content, and play a role in increasing the classification accuracy. A dedicated study of the influence of the distribution of the $\mathcal{F}$-statistic in the sky for astrophysical signals and detector artifacts will be addressed in a separate study.

\section{Conclusions}
\label{sec:conc}

We proved that the CNN can be successfully applied in the classification of {\tt TD-Fstat search} results, multidimensional vector distributions corresponding to three signal types: the GW signal, the stationary line and the noise. We compared 2D and 1D implementations of CNN. The latter achieved much higher accuracy (94\% with respect to 85\%) over candidate signals generated for broad range of frequencies and $\rho_{inj}$. For majority of signals ($\rho_{inj}{\geq}10$) 1D CNN maintained more than 90\% of accuracy. This level of accuracy was preserved at the classification of the signal candidates injected in bands of unknown frequency (i.e. we show that the constructed CNNs are able to generalize the context). 

2D CNN represented a different character. Although, the overall accuracy was worse than 1D model, the 2D version seemed to achieve better results as a binary classifier (between the {\bf cgw} and the {\bf noise}). Representation of the input data in the form of the image seemed to cause significant problems for the proper classification of the {\bf line}. Even though the 2D CNN had worse generalization ability, it was able to outperform the 1D implementation for the narrow-band frequencies 100 Hz and below. Nevertheless, 1D CNN with its ability to generalize unknown samples (in particular with respect to the frequency) seemed to be better choice for the realistic applications.

This project, as one of the few, researches the application of DL as a supplementary component to MF. Adopting signal candidates as the DL input instead of raw data allows to avoid problems that other researchers encountered. This approach limits the number of signals to those that exceeded the $\mathcal{F}$-statistic threshold, i.e. analysed distribution instances are firmly characterized by known significance. As Gebhard et al. \cite{2019arXiv190408693G}) described, application of DL on raw data provides signal candidates of unknown or hard to define significance. Before DL could be used as a safe alternative to MF for the detection of GW, it has to be studied further. However, our results can already be considered in terms of supporting role to MF. For example, it could be applied to the pre-processing of signal candidates for the further steps follow-up via fast classification, and to limit the parameter space to be processed further. As our results shows, a relatively simple CNN can also be used in the classification of spectral artifacts e. g., as an additional tool for flagging and possibly also removing spurious features from the data. Among the many possibilities for further development within the are of CW searches we are considering is also the application of DL in the follow-up of signal candidates in multiple data segments (post-processing searches for patterns), as well as the analysis of data from the network of detectors.

\section*{Acknowledgments}
\label{sec:ack}
The work was partially supported by the Polish National Science Centre grants no. 2016/22/E/ST9/00037 and 2017/26/M/ST9/00978, the European Cooperation in Science and Technology COST action G2Net no. CA17137, and the European Commission Framework Programme Horizon 2020 Research and Innovation action ASTERICS under grant agreement no. 653477. The Quadro P6000 used in this research was donated by the NVidia Corporation via the GPU seeding grant. This research was supported in part by PL-Grid Infrastructure 
(the productions runs were carried out on the ACK Cyfronet AGH Prometheus cluster). The authors thank Marek Cie{\'s}lar and Magdalena Sieniawska for useful insights and help with editing this publication.

\section*{Data Availability Statement}
The data that support the findings of this study are available from the corresponding author upon reasonable request.

\section*{References}
\bibliography{bibfile}

\end{document}